\documentstyle[12pt]{article}
\begin{document}

\begin{center}
{\Large  {\bf Possible physical manifestation of the Weyl
non-Abelian gauge field} \\} \vspace*{1cm} {\bf B.M.Barbashov,
A.B.Pestov\\} \vspace*{2cm}{\bf Joint Institute for Nuclear Research
\\ Bogoliubov Laboratory of Theoretical Physics, Dubna } \end{center}

\begin{abstract}
 On the basis of the  Weyl equations of congruent transference, we consider
a possible influence of the Weyl non-Abelian gauge field defining the
transference on the precession of a gyroscope.
Plane-wave solutions to the equations of the
non-Abelian gauge field are derived.
\end{abstract}
1.In ref.[1] it has been shown that the congruent transference introduced
by Weyl [2] in 1922 defines a non-Abelian gauge field. In
ref.[3], we have constructed a spinor representation of the Weyl gauge
group and found the corresponding spinor current that is a source of the
above gauge field. Since there exists the spinor source of a non-Abelian
gauge field, it makes sense to question about observable manifestations
of that sort of interactions on macroscopic scales. One possibility is
considered in this note. It is shown that if a rotating body undergoes
neither electromagnetic nor gravitational forces, the precession of the
intrinsic angular momentum can be caused by the non-Abelian gauge field
under consideration.

As it is known (see, e.g.,[4]), the most general law of parallel
transference of vectors along a given curve $x^i = x^{i}(\tau)$ is
defined by the system of ordinary differential equations of the form
$$ \frac{dS^i}{d\tau}= - \Gamma^i_{jk}u^j S^k ,$$ where $u^i=dx^i/
d\tau.$  From the condition for the length of a vector being
conserved in the parallel transference  it follows that the
coefficients of linear connection $\Gamma^i_{jk} $ should obey the system
of algebraic equations
$$\partial_j g_{ik} - \Gamma^l_{ji} g_{lk} -
\Gamma^l_{jk}g_{il} = 0.$$
It is assumed that the metric tensor
$g_{ij}$ is given, and the scalar product is defined by $g(S,S) =
g_{ij} S^i S^j.$ The general solution to this system of equations can be
written in the form [2]
\begin{equation} \Gamma^i_{jk} = \{^i_{jk}\} +
G^i_{jk}, \end{equation} where $G^i_{jk} = g^{il} G_{jkl}$ and $G_{jkl}$
is a second-rank tensor field skew-symmetric in the last two indices
$G_{jkl} + G_{jlk} = 0.$ It is just the Weyl non-Abelian gauge field
whose properties have been considered in the above-mentioned papers.
In formula (1),  $\{^i_{jk} \}$ denote, as usual, the components
of  Levi-Civita connection  of the metrics $g_{ij}$ $$
\{^i_{jk} \} = \frac{1}{2}g^{il} (\partial_j g_{kl} + \partial_k
g_{jl} - \partial_l g_{jk}).$$ Connection (1) defines the congruent
transference introduced by Weyl according to which a vector field is
changed by that transference as follows: \begin{equation}
\frac{dS^i}{d\tau} + \{^i_{jk} \}u^j S^k = - G^i_{jk}u^j S^k.
\end{equation} The law of transference (2) includes the displacement
belonging to the Riemannian geometry and rotation defined by the
metric $g_{ij}$ and bivector $G_{jkl}u^j.$ The Weyl gauge theory is a
realization of the abstract theory of gauge fields within the
differential geometry whose specific feature is that it does not
distinguish between space-time and a gauge space [1]. In this note,
we discuss a possible physical manifestation of the non-Abelian gauge
field $G^i_{jk}$.

2. Let us consider a rotating body, for instance, a gyroscope, that
is accelerated by forces applied to its center of mass. Forces of
this sort produce no torque, therefore, they do not change the vector
length $S^i$ of the proper angular momentum, but appear as its
precession called the Thomas precession. We will apply equations (2)
to describe the Thomas precession of a gyroscope initiated by the
Weyl gauge field. The produced rotation should preserve the
orthogonality of the vector of the intrinsic angular momentum of a
body $S^i$ and 4-velocity $u^i.$  In view of (2) and the gauge tensor
field $G_{ijk}$ being skew-symmetric in the last two indices, we
obtain $$\frac{\delta }{\delta \tau} (u_i S^i) = \frac{\delta
u^i}{\delta \tau} S_i + u_i \frac{\delta S^i}{\delta \tau} =
\frac{\delta u^i}{\delta \tau} S_i - u_i G^i_{jk}u^j S^k =
(\frac{\delta u^i}{\delta \tau} + G^i_{jk} u^j u^k ) S_i,$$
where $\delta / \delta \tau $ is an absolute derivative. Then it follows that
if the center of gravity of a gyroscope moves along the trajectory
that is given by the equations of geodesics in connection (1)
\begin{equation} \frac{\delta u^i}{\delta \tau}  + G^i_{jk}u^j u^k
=0, \end{equation}
then
$$ \frac{\delta }{\delta \tau} (u_i S^i) = 0.  $$
Thus, equations (2) are capable of describing precession of the
proper angular momentum of a body in the framework of
dynamics given by equations (3). Let us discuss them in greater
detail.

3. Of great importance are quantities irreducible with respect to
the fundamental group of symmetry. If this principle is applied to
the field $G_{ijk},$, it is necessary to consider global transformations
of the Weyl gauge group that is structured like the Lorentz group.
The tensor $G_{ijk},$ with respect to this group is decomposable into
irreducible components. According to  the results of ref. [5] this
expansions looks as follows:
\begin{equation} G_{ijk}= \frac{2}{3}(T_{ijk} - T_{ikj}) +
\frac{1}{3}(g_{ij} F_k - g_{ik} F_j) + e_{ijkl}A^l, \end{equation}
where $e_{ijkl} $ is a completely antisymmetric Levi-Civita tensor and
other quantities are defined in the following way:

$$ F_i = g^{jk}G_{jki}, \quad A_i =
\frac{1}{6}e_{ijkl}G^{jkl}, $$   \\
$$T_{ijk}
=\frac{1}{2}(G_{ijk} + G_{jik}) + \frac{1}{6}(g_{ki}F_j + g_{kj}F_i )
-\frac{1}{3} g_{ij} F_k.$$

Consider the case when irreducible components $T_{ijk} = 0,
A_i = 0.$ Then it can be set
\begin{equation} G_{ijk}= g_{ij} F_k -
g_{ik} F_j .\end{equation}
Inserting (5) into (3), we get
\begin{equation} \frac{\delta u^i}{\delta \tau}  + F^i - u^i (u_k
F^k) =0, \end{equation} since $u_i u^i = 1.$
Within the relativistic mechanics [6],
an accelerated motion of a particle  with mass $m$ is determined by equations
\begin{equation}a^i = \frac{\delta u^i}{\delta \tau} = \frac{d^2
x^i}{d\tau^2} + \{^i_{jk} \}\frac{dx^j}{d\tau} \frac{dx^k}{d\tau} =
\frac{1}{m} N^i, \end{equation} where $N^i$ is a 4-force orthogonal to
a 4-velocity. So, an irreducible component given by the polar vector
$F^i,$ permits physical interpretation in the framework of relativistic
mechanics provided that $\frac{1}{m} N^i = -F^i
 + u^i (u_k F^k).$

Consider now how substitution (5) is consistent
with the equations of motion of the field $G_{ijk},$ proposed in
[1],\,[3].  Equations of the gauge field $G_{ijk}$ are of the form
\begin{equation} \nabla_i B^{ijkl} + G^k_{im} B^{ijml} - G^l_{im}
B^{ijmk} + S^{jkl}= 0,\end{equation} where \begin{equation}
B_{ijkl} = \nabla_i G_{jkl} - \nabla_j G_{ikl} + G_{ikm} G^m_{jl}
-G_{jkm} G^m_{il} + R_{ijkl}\end{equation} is the strength tensor
and the tensor $S^{jkl}$ is the current, a source of the gauge field under
consideration. It is assumed that the current $S^{ijk}$ is given and obeys
 the equations $$ \nabla_i S^{ijk} + G^j_{im} S^{imk} -
 G^k_{im} S^{imj} = 0  $$ that are analogous to the local conservation law
 of energy-momentum in General relativity [3].

Inserting (5) into (9), we get \newpage $$ B_{ijkl}=
$$ $$g_{ijkl} (F_m F^m) + g_{ijmk} (\nabla^m + F^m ) F_l - g_{ijml}
(\nabla^m + F^m ) F_k + R_{ijkl}, $$ where $g_{ijkl} = g_{ik} g_{jl} -
g_{il} g_{jk}.$ The system of equations (5) and (8) for the field
$F_i$ is complicated and, besides, the number of equations is much larger
than the number of unknown functions. Therefore, one should
verify whether system (5) and (8) is consistent or not. Contracting
equations (8) in indices $j$ and $l,$ we arrive at the equations
 \begin{equation} \nabla_i B^{ik} + G^k_{im} B^{im} + G_{ijm} B^{ijmk} +
 S^{k}= 0,\end{equation} where $B^{ik} = g_{jl}B^{ijkl},$ $S^k =
g_{jl}S^{jkl}.$ Contraction in indices $j$ and $k,$ produces analogous
equations; whereas that in $k$ and $l$, an identity. So, any solution to
eqs. (8) is a solution to eqs. (10). Inserting (5) into
(10), we get
 \begin{equation} \nabla_i B^{ik} - F_m( B^{mk} + B^{km} ) + F^k B
+ S^k =0 , \end{equation} where $B = g_{ij} B^{ij}.$ Like the
 strength tensor $B_{ijkl} $  we can express the
 tensors $B_{ij}, \, B$ in terms of the field  $F_i$:
  \begin{equation} B_{ij} = 2g_{ij} F_m F^m - 2 (\nabla_i + F_i) F_j
  - g_{ij}\nabla_m F^m + R_{ij}, \end{equation} \begin{equation} B =
  6 (F_m F^m - \nabla_m F^m) + R, \end{equation} where $R_{ij}$ is
  the Ricci tensor, the $R$-scalar curvature of space-time.
 Substituting (12) and (13) into (11), we find that any solution to
equations (5) and (8) is a solution to the equations \begin{equation}
\nabla^i \nabla_i F^k + 3 F^k (\nabla_i F^i - F_i F^i) +
\frac{1}{2}\nabla^k (\nabla_i F^i - 3 F_i F^i ) = \frac{1}{2}S^k ,
\end{equation} where we put $R_{ijkl} = 0.$

To prove the consistency of system of equations (5) and (8) we show that
it possesses nontrivial solutions at $S^{ijk} = 0, R_{ijkl} =0.$
One of the solutions looks as follows:
$$F_i = \frac{p_i}{p_0 x^0 + p_1 x^1 +
  p_2 x^2 + p_3 x^3} = \frac{p_i}{p_m x^m},$$ where $p_i$ is a constant
non-isotropic vector field, $p_i p^i \neq 0.$

It can be verified that eqs. (14) are fulfilled with that solution.
Calculating the strength tensor we obtain
  $$B_{ijkl} = g_{ijkl} \frac{p_m p^m}{(p_n x^n)^2}.$$
and inserting this expression into field equations (8) we see that they
also hold valid.

 Another solution to eqs. (5) and (8) without sources is a plane wave of
an arbitrary shape: $$F_i = k_i \Phi(kx),$$ where
 $k_i$ is an isotropic (a null) vector, $k_i k^i =0,$ $ kx = k_i x^i.$
In this case, the strength tensor is given by
$$B_{ijnl} = (\Phi^{'}+
 \Phi^2) (g_{ijmn} k_l - g_{ijml}  k_n)k^m ,$$ where $\Phi^{'}$ is the
derivative with respect to argument. Thus, system of equations (5) and (8)
is consistent.

If in equations (14) we omit the terms with self-action and put
 $S^k = 0,$ we obtain the equations which are a particular case (at $\lambda
 =3/2$) of the equations
 $$\nabla_i \nabla^i A^k - (1-\lambda)
\nabla^k \nabla_i A^i = 0,$$
 that are employed in quantum electrodynamics for fixing gauge of the
 vector potential $A^i$ [7].

4. To summarize, we dwell on a possible physical interpretation of our
consideration. If a rotating body which does not undergo any torque but
suffers acceleration, the direction of its rotation with respect to the
inertial system changes in accordance with the rule of transference
(2). Then it follows that we can qualitatively try to discover the Weyl
non-Abelian gauge field by observing the behaviour of a gyroscope at
a cosmic station. Since gravitational and electromagnetic forces can
there be assumed zero, the gyroscope precession, if discovered, could
be induced by that gauge field.

We express our gratitude to R.Asanov and E.Kuraev for useful discussions
and to I.Tyutin for constructive remarks. The work was partly supported by
the Russian Foundation for Basic Research, grant N 07-01-00745.

\end{document}